\documentclass[a4paper,11pt]{article}
\usepackage{graphicx}
\usepackage{xcolor}
\usepackage{amsmath,amsfonts,amssymb,amstext,graphicx,hyperref,hypcap}
\usepackage{cancel}
\usepackage{empheq}
\usepackage{float}
\usepackage{cite}
\usepackage{subcaption}
\usepackage{authblk}
\usepackage{comment}
\usepackage{blindtext}
\usepackage{cases}
\usepackage{ulem}
\usepackage{epsfig}%
\usepackage[bottom]{footmisc}

\usepackage{epsfig}%
\usepackage[bottom]{footmisc}
\begin{document}
\date{}

\title{\centerline \bf Stability analysis for a cosmologically viable model of $f(R)$ gravity}
\bigskip

\author{Sayantan Dey\thanks{sayantand84@gmail.com}} 
\affil{Ramakrishna Mission Vivekananda Educational and Research Institute, 
Belur Math, Howrah 711202, India} 
\maketitle

\begin{abstract} 
In recent times, there has been an increasing interest with theories of modified gravity as a
means to gain a deeper understanding of the universe's late-time acceleration phase. 
In this study we   focused our attention on
a specific cosmologically viable $f(R)$ model.
We performed a dynamic stability analysis of this model, revealing that
the model supports presence of just one asymptotically stable solution which
can explain the present-day acceleration of the universe.  
\end{abstract}

\section{Introduction}
\label{sec:1}
Cosmological observations over last few decades
have revealed that the present universe is undergoing an accelerated expansion and a transition from a decelerating phase to the current accelerated phase occurred during the late time phase
of cosmic evolution. Observation of luminosity distance and redshift measurements of Type Ia supernovae (SNe Ia) events \cite{r1,r2,r3,r4}
provided the initial evidences in favour of these observations.
These observations receive additional confirmation through the detection of baryon acoustic oscillations  \cite{r5,r6}, the study of cosmic microwave background radiation\cite{r7,r8,r9}, and the analysis of the power spectrum of matter distributions throughout the universe \cite{r10,r11}.  The commonly employed term to characterize the root cause behind the observed cosmic acceleration during late-time is `dark energy' which postulated as a hypothetical form of unclustered energy  that exerts a negative pressure,  opposing thereby the gravitational attraction, ultimately leading to the cosmic acceleration.\\

In addition to the observed late-time cosmic acceleration, there is also the theoretical concept of cosmic inflation, which posits a rapid and exponential expansion of the universe during its initial stages. While direct observational confirmation of cosmic inflation is lacking, this idea is highly compelling as it offers solutions to various cosmological enigmas and generates consistent predictions across multiple observations. These include phenomena such as the Cosmic Microwave Background Radiation, the large scale structure of the universe, the apparent near-flatness of the present-day universe, and the potential existence of primordial gravitational waves.\\

Einstein's field equation, which is a cornerstone of the general theory of relativity, describe how the presence of matter and energy in the cosmos shapes the geometry of spacetime. These equations have proven to be exceptionally effective in elucidating gravitational interactions across a wide spectrum of scenarios, encompassing everyday experiences, as well as celestial observations and experimental contexts.\\

When Einstein's equations are employed to describe the universe at large scales with a homogeneous and isotropic 
spacetime geometry described by the Friedmann-Robertson-Walker (FRW) metric,  they yield the Friedmann equations of form:
\begin{eqnarray}
\left(\frac{\dot{a}}{a}\right)^2
&=& \frac{8\pi G}{3}\rho - \frac{K}{a^2}
\label{eq:a1}\\
\left(\frac{\ddot{a}}{a}\right)
&=& - \frac{4\pi G}{3} (\rho + 3p)
\label{eq:a2}\,,
\end{eqnarray}
where the curvature constant $K$ and time-dependent scale factor $a(t)$ that accommodates 
the observed expansion of the universe characterize the FRW metric and
the content of the universe being  considered to be
an ideal fluid with energy density  $\rho$ and  pressure  $p$. $G$ is the Newton's gravitational constant. 
With the matter and radiation components of the universe, the expression $(\rho+3p)$ in Eq.\ (\ref{eq:a2}) is
always   positive, leading to the conclusion that $\ddot{a}$ is invariably negative throughout the entire cosmic timeline. Consequently, according to the Einstein equation, this results in a decelerating expansion of the universe, thereby failing to generate the desired acceleration during both the late-time evolution phase and the inflationary era.\\

Numerous theoretical approaches, emerging from diverse perspectives, have been developed to construct models aimed at elucidating the phase of accelerated cosmic expansion. One notable model in this regard is the $\Lambda$-CDM model \cite{r12}, which closely aligns with contemporary cosmological observations but grapples with the fine-tuning problem when scrutinized through the lens of particle physics.
Furthermore, alternative field-theoretic models have been put forth, involving the introduction of (scalar) fields as constituents of the universe, resulting in the generation of specific forms of the energy-momentum tensor featuring negative pressure within Einstein's equations. These models include quintessence \cite{r13} and k-essence \cite{r14}, wherein the driving mechanisms behind cosmic acceleration are scalar fields characterized by slowly evolving potentials and scalar field kinetic energy, respectively.\\

Another vital class of models, referred to as modified gravity models, seek to explain the cosmic acceleration by manipulating the geometry itself, without making alterations to the energy-momentum tensor within Einstein's equations. These modifications predominantly relate to the geometric aspect of Einstein's equation, often originating from higher-order corrections within the Einstein-Hilbert action. By making these tailored adjustments, it becomes feasible to induce cosmic acceleration.
The simplest form of modification entails substituting the Ricci scalar $R$ in Einstein's equations with an arbitrary function of $R$ denoted as $f(R)$. The meticulous selection and justification of this specific arbitrary function play a pivotal role in all modified gravity models. Theoretical considerations, including the requirement for a theory that avoids various instabilities while maintaining stable perturbations, as well as its ability to accurately replicate observed cosmic evolutionary patterns and the behaviour of local systems, serve as the primary means for  constraining these $f(R)$ models.\\

Obtaining analytical solutions for $f(R)$ theories in the context of modified gravity is an exceedingly difficult task due to the presence of nonlinear terms in the field equations. To address this challenge, one can employ dynamical system analysis \cite{r15}
as an alternative approach. In the field of cosmology, dynamical analysis plays on a crucial role in the investigation $f(R)$ gravity theories. This analytical method is takes care of unique degrees of freedom introduced by the chosen functional form of $f(R)$, 
eventually assessing the stability of cosmological solutions within these theories, thus ensuring the emergence of physically plausible cosmic scenarios. Furthermore, dynamical analysis has the capacity  constrain the parameter space associated with $f(R)$ models, aligning them with empirical evidence from observed cosmological data.\\

There exist two different approaches  for describing the gravitational field's behavior and its interaction with matter within the framework of Einstein's theory of general relativity. The more commonly utilized approach, known as the metric formalism, treats the metric tensor, denoted as $g_{\mu\nu}$, and the connections represented by the Christoffel symbols as a unified entity. Both of these components are determined as solutions to the same set of equations. In contrast, the Palatini formalism treats the metric and the connection as independent variables and varies them separately to derive equations of motion. This results in distinct field equations for the metric and the connection.\\

For the purposes of this study, we exclusively consider the metric formalism. Within this formalism, a selection of $f(R)$ models has been examined, all of which satisfy the cosmological viability conditions (as detailed in references \cite{r16} and \cite{r17}). The chosen list of such $f(R)$ models includes notable ones like the Hu \& Sawicki model  \cite{r18}, the Starobinsky model \cite{r19}, the Tsujikawa model \cite{r20}, the exponential model \cite{r21}, arcTanh models \cite{r22}, among others. In our investigation, we focus on a specific model defined by the equation:
\begin{eqnarray}
f(R) &=& R - \alpha R_C (R/R_C)^n
\label{eq:a3}
\end{eqnarray}
Here, the model parameters must satisfy the conditions: $0 < n < 1$,  $\alpha, R_C >0$   to ensure the model's coslomogical viability, as discussed and considered in reference \cite{r17}. Our study involves a dynamical analysis of this model, where we examine the significance of various fixed points within the scenario.\\

In Sec.\ \ref{sec:2} we present a brief outline of 
the theoretical framework of the 
$f(R)$-gravity and in Sec.\ \ref{sec:3}, we provide 
the basics of the dynamical tenchnique and the results
of the dynamical analysis for the model in Eq.\ (\ref{eq:a3}).
We present the conclusions in Sec.\ \ref{sec:4}

\section{A Brief outline of the framework of $f(R)$-Gravity}
\label{sec:2}
The generalized Einstein-Hilbert action in the context of $f(R)$ theories can be expressed as follows:
\begin{eqnarray}
S &=& \dfrac{1}{2\kappa^{2}}\int d^{4}x\sqrt{-g} f(R) + S_{m}\,.
\label{eq:a4}
\end{eqnarray}
Here $\kappa^{2} = 8\pi G$, $g$ represents the determinant of the metric tensor, $S_m$ stands for the action corresponding to the matter (and radiation) field. Variation
of the above action with respect to the metric tensor gives 
the field equation in metric formalism as
\begin{eqnarray}
F(R)R_{\mu\nu}-\frac{1}{2}f(R)g_{\mu\nu}-\nabla_{\mu}\nabla_{\nu} F(R)+g_{\mu\nu}\Box  F(R) &=& k^{2} T_{\mu\nu}
\label{eq:a5}
\end{eqnarray}
where $ F(R)= df/dR \equiv f_{,R}$ , $\Box = \nabla_{\mu}\nabla^{\mu}$ and $T_{\mu\nu}$ is given by 
\begin{eqnarray}
T_{\mu\nu} &=& \dfrac{-2}{\sqrt{-g}} \dfrac{\partial S_{m}}{\partial g^{\mu\nu}}
\label{eq:a6}
\end{eqnarray}
Taking trace on both sides of the above equation we get 
\begin{eqnarray}
3\Box F(R)+F(R)R-2f(R)=k^{2}T
\label{eq:a7}
\end{eqnarray}
where $T=g^{\mu\nu}T_{\mu\nu}$. It's worth noting that in this context, the relationship between the Ricci scalar $R$, and the stress-energy tensor $T$, is differential rather than algebraic, as seen in General Relativity, where we have the equation $R+k^{2}T=0$. This distinction suggests that the field equations of $f(R)$ theories will allow for a broader range of solutions compared to Einstein's theory. \\

As discussed in \cite{r16,r17}, the conditions that has to
be satisfied by any viable $f(R)$ model are listed below along
with the corresponding considerations from which they conditions
would arise.
\begin{center}
\begin{tabular}{lcl}
$f_{,R} >0$ for $R>R_0$ &:& (to avoid anti-gravity)\\
$f_{,RR} >0$ for $R>R_0$ &:& (consistency with local gravity tests)\\
$f(R) \approx R - 2\Lambda$ for $R \gg R_0$ &:& (consistency with local gravity tests and\\
&& \quad existence of matter dominated epoch)\\
$0 < Rf_{,RR} / f_{r} < 1$ at $r=-Rf_{,R}/f = -2$
& :& (stability of the
late-time de-Sitter solution)
\end{tabular}
\end{center}
where, $R_0$ in the above expressions is the 
value of the Ricci scalar at present epoch.
We may verify the $f(R)$ model in Eq.\ (\ref{eq:a3}),
which we choose to investigate here, satisfies all the above conditions
for the model parameters $0<n<1$, $\alpha, R_C>0$.\\

We considered the universe to be isotropic and
homogeneous at large scale and described by spatially flat
FRW metric (i.e. with $K=0$):
\begin{eqnarray}
ds^2 &=& -dt^2 + a^2(t) \Big{[} dr^2 + r^2(d\theta^2 + \sin^2\theta d\phi^2)\Big{]}
\label{eq:a8}
\end{eqnarray}
where $t$ is the cosmic time, $a(t)$ is the scale factor
and $(r,\theta,\phi)$ represent the comoving coordinates.
For this flat FRW spacetime geometry the Ricci scalar $R$
is related to the Hubble parameter $H \equiv \dot{a}/a$ by
the relation
\begin{eqnarray}
R &=& 6(2H^2 + \dot{H})
\label{eq:a9}
\end{eqnarray}
In the framework of the FLRW universe, which is filled with an ideal perfect fluid characterized by energy density $\rho$ and pressure $p$, the modified field equations (\ref{eq:a5}) can be expressed as follows for the `00' and `$ii$' components:
\begin{eqnarray}
3FH^{2} &=& \kappa^{2}\rho + \dfrac{1}{2}(FR-f)-3H\dot{F}
\label{eq:a10}\\
 -2F\dot{H} &=& k^{2}(\rho + p)+\ddot{F}-H\dot{F}
\label{eq:a11}
\end{eqnarray}
We can further decompose the energy density $\rho$ as the sum of two components: $\rho = \rho_m + \rho_r$, where $\rho_m$ represents the energy density due to matter, and $\rho_r$ represents the energy density due to radiation. Additionally, assuming that the matter component is nonrelativistic (similar to dust), the pressure $p_m$ due to matter is negligible ($p_m \sim 0$). Therefore, the total pressure of the fluid is solely due to radiation and is equal to one-third of the radiation energy density, expressed as $p = \rho_r/3$. Under such considerations, Eqs.\ (\ref{eq:a10})
and (\ref{eq:a11}) takes the following forms:
\begin{eqnarray}
 3FH^{2} &=& \kappa^{2}(\rho_m + \rho_r) + \dfrac{1}{2}(FR-f)-3H\dot{F}
\label{eq:a12}\\
 -2F\dot{H} &=& k^{2}( 4 \rho_r /3 + \rho_m  )+\ddot{F}-H\dot{F}
\label{eq:a13} 
\end{eqnarray}
Note that, 
we can alternatively represent Eq. (\ref{eq:a12}) using density parameters associated with matter, radiation, and geometric curvature (GC) within the framework of the modified gravity theory as
\begin{eqnarray}
\Omega_m + \Omega_r + \Omega_{\rm GC} &=& 1
\label{eq:a13a}
\end{eqnarray}
where
\begin{eqnarray}
\Omega_{m} \equiv \dfrac{k^{2}\rho_{m}}{3FH^{2}}\,,\quad
\Omega_{r} \equiv \dfrac{k^{2}\rho_{r}}{3FH^{2}} \,,\quad
\Omega_{\rm GC} \equiv \dfrac{1}{3FH^{2}}
\left[\dfrac{1}{2}(FR-f)-3H\dot{F}\right]
\label{eq:13b}
\end{eqnarray}

\section{Results of Dynamical analysis}
\label{sec:3}
Following the approach as in \cite{r17}, we define the dimensionless variables 
\begin{eqnarray}
 x_{1}\equiv -\dfrac{\dot {F}}{HF}\,,\quad       x_{2}\equiv -\dfrac{f}{6FH^{2}}\,,\quad     x_{3}\equiv \dfrac{R}{6H^{2}}=\dfrac{\dot{H}}{H^{2}}+2\,,\quad   x_{4}\equiv \dfrac{k^{2}\rho_{r}}{3FH^{2}}
\label{eq:a14}
\end{eqnarray}
Expressing the density parameters associated with various components in terms of the dimensionless quantities $x_i$, we can write 
\begin{eqnarray}
\Omega_{m}  = 1 -x_{1}-x_{2}-x_{3}-x_{4} \quad , \quad
\Omega_{r}  = x_{4}
\quad \mbox{and } \quad \Omega_{\rm GC} = x_1 + x_2 + x_3
\label{eq:a15}
\end{eqnarray}
By introducing a dimensionless parameter, $N = \ln a$, to account for temporal variations, 
we can represent the
Eqs.\ (\ref{eq:a12}) and Eqs.\ (\ref{eq:a13}) as the following
set of autonomous equations:
\begin{eqnarray}
y_1 \equiv \dfrac{dx_{1}}{dN} &=& -1-x_{3}-3x_{2}+x_{1}^{2}-x_{1}x_{3}+x_{4} \label{eq:a16}\\
y_2 \equiv \dfrac{dx_{2}}{dN} &=& \dfrac{x_{1}x_{3}}{m}-x_{2}(2x_{3}-4-x_{1}) \label{eq:a17}\\
y_3 \equiv \dfrac{dx_{3}}{dN}&=&-\dfrac{x_{1}x_{3}}{m}-2x_{3}(x_{3}-2)\label{eq:a18}\\
y_4 \equiv \dfrac{dx_{4}}{dN}&=&-2x_{3}x_{4}+x_{1}x_{3} \label{eq:a19}
\end{eqnarray}
where 
\begin{eqnarray}
m\equiv\dfrac{d \ln F}{d\ln R} = \dfrac{Rf_{,RR}}{f_{,R}}
\quad \mbox{and }   
r\equiv\dfrac{d \ln f}{d\ln R}=-\dfrac{Rf_{,R}}{f}=\dfrac{x_{3}}{x_{2}}
\label{eq:a20}
\end{eqnarray}
The effective equation of state $\omega_{\rm eff}$ of the 
fluid content of the universe can then be expressed as
\begin{eqnarray}
\omega_{\rm eff}
& \equiv & 
\dfrac{p}{\rho} = \dfrac{\rho_r/3}{\rho_m + \rho_r}
= -1 - \frac{2}{3}\dfrac{\dot{H}}{H^2}
= - \frac{1}{3}(2x_3 - 1)
\label{eq:a21}
\end{eqnarray}
Given that both $m$ and $r$ are functions of $R$, it is possible, in principle, to consider $m$ as a function of $r$, denoted as $m = m(r)$.

The fixed points of the system correspond to
the solutions $y_i \equiv \dfrac{\dot{x}_i}{N} = 0$.
Through a linear stability analysis, we can determine the nature of these fixed points by examining the eigenvalues of the Jacobian matrix
$J = \Big{\vert}\Big{\vert}\dfrac{\partial y_i}{\partial x_j}\Big{\vert}\Big{\vert}$ evaluated at the fixed points.
The fixed points can be categorized as either hyperbolic or non-hyperbolic. If all the eigenvalues of the Jacobian matrix $J$ have a non-zero real part at any fixed point, it is classified as a hyperbolic fixed point. Conversely, if they do not have a non-zero real part, it is termed a non-hyperbolic fixed point.
For non-hyperbolic fixed points, linear stability theory is insufficient, and alternative methods such as Lyapunov's method or Central manifold theory must be employed to determine their nature.
However, in the case of a hyperbolic fixed point, the nature of the fixed point can be inferred based on the signs of the real parts of its eigenvalues. If all the eigenvalues possess a positive real part, the corresponding fixed point is unstable and acts as a repeller. On the other hand, if all the eigenvalues have a negative real part, it is considered a stable fixed point or attractor. When at least a pair of eigenvalues have real parts with opposite signs, the fixed point is classified as a saddle point.\\

We now employ the previously outlined analytical approach to examine the $f(R)$ model specified in Eq.\ (\ref{eq:a3}): $f(R) =  R - \alpha R_C (R/R_C)^n$, with $0<n<1$, $\alpha, R_C>0$.
Our primary focus is on the late-stage cosmic evolution, during which the energy density attributed to the radiation component becomes negligible. Consequently, we set $x_4=0$ in the autonomous equations (\ref{eq:a16}), (\ref{eq:a17}), and (\ref{eq:a18}), while excluding Equation (\ref{eq:a19}). This results in a set of three autonomous equations for the system, as follows:
\begin{eqnarray}
 \dfrac{dx_{1}}{dN} &=& -1-x_{3}-3x_{2}+x_{1}^{2}-x_{1}x_{3}  \label{eq:a22}\\
 \dfrac{dx_{2}}{dN} &=& \dfrac{x_{1}x_{3}}{m}-x_{2}(2x_{3}-4-x_{1}) \label{eq:a23}\\
 \dfrac{dx_{3}}{dN}&=&-\dfrac{x_{1}x_{3}}{m}-2x_{3}(x_{3}-2)\label{eq:a24} 
\end{eqnarray}
Additionally, for the considered $f(R)$ model, we have the following relationships:
\begin{eqnarray}
r = \dfrac{x_{3}}{x_{2}} \quad \mbox{and } \quad m(r) = n\left(\dfrac{r+1}{r}\right) = n\left(\dfrac{x_{2}+x_{3}}{x_{3}}\right)
\label{eq:a25}
\end{eqnarray}

This autonomous system described by Eqs.\ (\ref{eq:a22},\ref{eq:a23},\ref{eq:a24})
exhibits four stationary points, as detailed in Table \ref{tab:1}. We have also provided the corresponding values of $\omega_{\rm eff}$ and $\Omega_m$ at these fixed points. In the following analysis, we investigate the inherent characteristics of each of these fixed points.
\begin{table}[h]
\centering
\begin{tabular}{ccccc}
&& $(x_1,x_2,x_3)$ & $\omega_{\rm eff} $ & $\Omega_m $\\
\hline
$P_1$ &:& $(-4,5,0)$  & $1/3$ & $0$ \\
$P_2$ &:& $(0,-1,2)$  & $-1$ & $0$ \\
$P_3$ &:& $\left(\dfrac{3n-3}{n}, \dfrac{3-4n}{2n^{2}}, \dfrac{4n-3}{2n}\right)$ & $\dfrac{1-n}{n}$ & $ \dfrac{-3+13n-8n^{2}}{2n^{2}}$\\
$P_4$ &:& $\left(\dfrac{2 (n-2)}{2 n-1},\dfrac{5-4 n}{2 n^2-3 n+1},\dfrac{n (4 n-5)}{2 n^2-3 n+1}\right)$ & $-\dfrac{1}{3}\left(\dfrac{6n^{2}-7n-1}{2n^{2}-3n+1}\right)$ & $0$\\
\hline
\end{tabular}
\caption{Fixed points of the autonomous system described by Eqs.\ (\ref{eq:a22},\ref{eq:a23},\ref{eq:a24}) and the values of $\omega_{\rm eff} $ and  $\Omega_m $ at the fixed points.}
\label{tab:1}
\end{table}

For the purpose of conducting a stability analysis, we calculate the eigenvalues of to the Jacobian matrix
 $J= \Big{\vert}\Big{\vert}\dfrac{\partial y_i}{\partial x_j}\Big{\vert}\Big{\vert}$. It's important to note that this matrix is of dimension $3\times 3$ in the context of our current study since we have   excluded the radiation component during the current epoch. Below, we elaborate on the pertinent findings of this analysis. \\

\begin{description}
\item 
The eigenvalues of $J$ at the fixed point $P_1$ are $-5,4,-3$ each of which are independent
of the model parameters. Since a pair of eigenvalues have
opposite sign, the $P_1$ is a saddle point. 

\item The eigenvalues corresponding to the point $P_2$ are 
\begin{eqnarray*}
\lambda_1 = -3\,,\quad 
\lambda_2 = -\dfrac{3}{2} -\dfrac{i}{2\sqrt{n}}\sqrt{32 - 25n}\,,\quad  
\lambda_3 =  -\dfrac{3}{2} +\dfrac{i}{2\sqrt{n}}\sqrt{32 - 25n} \,,\quad  
\end{eqnarray*}
Given that the parameter $n$ falls within the range $0 < n < 1$, it's important to note that the term $(32 - 25n)$ is consistently positive. As a result, the   eigenvalues $\lambda_2$ and $\lambda_3$ adopt the form of $(-3/2 \mp i\beta)$, where $\beta \equiv \dfrac{\sqrt{32 - 25n}}{2\sqrt{n}} \neq 0$ is real.
All three eigenvalues possess a negative real part, establishing the point $P_2$ as a stable attractor. This stability is further enhanced by the fact that two of the eigenvalues form a complex conjugate pair, imparting a spiral stability to point $P_2$.
 Furthermore,  at this stable attractor point $P_2$, 
 the effective equation of state parameter $\omega_{\rm eff} = -1$
 and the density
parameter due to geometric curvature $\Omega_{\rm GC} = 1$ ($\Omega_m = 0$)
Consequently, the stable attractor point $P_2$ corresponds to a phase of cosmic acceleration, with this acceleration being  driven by the geometric curvature term arising from the $f(R)$ modification.  

\item The eigenvalues of $J$ corresponding to the point $P_3$ are obtained as
\begin{eqnarray*}
\lambda_1 = \dfrac{3 \left(n^3-2 n^2+n\right)}{(n-1) n^2}\,,\quad 
\lambda_2 = \dfrac{-3 n^2 + 3 n -n\sqrt{Q}}{4 (n-1) n^2}\,,\quad 
\lambda_3 = \frac{-3 n^2 + 3 n + n\sqrt{Q} }{4 (n-1) n^2}
\end{eqnarray*}
where,  $Q=256 n^4-864 n^3+1025 n^2-498 n+81$.
The eigenvalues are dependent upon the model parameter $n$, thereby dictating the stability characteristics of point $P_3$. Additionally, the values of $\Omega_m$ (including $\Omega_{\rm GC}$) and $\omega_{\rm eff}$ are subject to the influence of the parameter $n$, as presented in  Tab.\ \ref{tab:1}.
 We conducted a thorough scan across the entire range of $0 < n < 1$ and determined that, within the interval $0.33 < n < 0.71$, all eigenvalues exhibit a negative real component. Notably, the   eigenvalues 
 $\lambda_2$ and $\lambda_3$ adopt the form $\delta \mp i\beta$, where $\delta$ is a real number and $\beta$ is a nonzero real value. Consequently, for the considered $f(R)$ model with $0.33 < n < 0.71$, point $P_3$ demonstrates spiral stability. Within this range of $n$ values, the effective equation of state parameter, $\omega_{\rm eff} = \frac{1-n}{n}$, maintains positive values within the range of $[0.42 - 2]$.
However, it is crucial to acknowledge that the constraint on the density parameter $\Omega_m = \dfrac{-3+13n-8n^{2}}{2n^{2}}$, which stipulates $0 < \Omega_m < 1$, permits a viable range of $n$ values only in the small window of [0.28-0.30]. This window falls outside the previously mentioned interval of $0.33 < n < 0.71$ where point $P_3$ exhibits spiral stability. Consequently, it is established that point $P_3$ lacks physical relevance in this context.

\item 
The eigenvalues of $J$ corresponding to the point $P_4$ are 
\begin{eqnarray*}
\lambda_1 =  -\frac{2 (n-2)}{2 n-1} \,,\quad 
\lambda_2 = \frac{-8 n^2+13 n-3}{2 n^2-3 n+1} \,,\quad 
\lambda_3 =  \frac{5-4 n}{n-1}
\end{eqnarray*}
All three eigenvalues are real, and upon scanning the entire range of $0 < n < 1$, it is observed that they all become negative within the interval $[0, 0.28]$ indicating that, in the specified range of
 $n$, point $P_4$ exhibits stability. This point   corresponds to $\Omega_{m}=0$ and
furthermore, based on the expression for the effective equation of state $\omega_{\rm eff} =-\dfrac{1}{3}\left(\dfrac{6n^{2}-7n-1}{2n^{2}-3n+1}\right)$ (see Tab.\ \ref{tab:1}), we obtained that $\omega_{\rm eff}$ takes on negative values for $0.5 < n \leqslant 1$. Notably, within this range, the negative value of $\omega_{\rm eff}$ is consistently more negative than -1, implying a phantom-like solution for cosmic acceleration. However, it's important to note that this phantom solution, occurring for $0.5 < n \leqslant 1$, is not a stable fixed point as this range of values for parameter $n$ excludes the interval $0 < n < 0.28$ in which point $P_4$ exhibits stability.
\end{description}

\section{Conclusion}
\label{sec:4}
In this article we have explored a specific cosmologically viable model of $f(R)$ gravity and investigated its potential to generate cosmic acceleration during the late time phase of cosmic evolution - an observationally established feature concluded from measurement of luminosity distance and redshifts of supernova events. 
We consider universe to be isotropic and homogeneous at large scales which is  described by a FRW spacetime metric. Within this framework, we represent the universe's constituents as a perfect fluid, defined by its energy density and pressure. The universe's evolutionary dynamics are postulated to be driven by a modified version of Einstein's field equations, where the modification involves substituting the Ricci scalar $R$ in Einstein's equation with a function of $R$ denoted as $f(R)$. One main reason for exploring $f(R)$ gravity models is the inability of Einstein's field equations in original form  to induce cosmic acceleration with matter (baryonic and dark) and radiation as its constituents. This prompts an investigation into how the function $f(R)$ can be leveraged to achieve this acceleration.
In order for a specific form (model) of $f(R)$ to be cosmologically viable, it is essential that it can  reproduce local gravitational phenomena and avoid anti-gravity effects. Furthermore, it should correctly predict the existence of a matter-dominated era during the cosmic evolution phase and demonstrate the stability of a late-time de-Sitter solution. 
We directed our attention to a specific $f(R)$ model, \textit{viz.} $f(R) = R - \alpha R_C (R/R_C)^n$, which satisfies all the above viability criteria within the parameter constraints of $0 < n < 1$ and $\alpha, R_C > 0$. Our investigation aimed to determine the parameter values of $n$ for which this model can induce a stable acceleration during late time phase of cosmic evolution. \\

We employ linear stability theory as our methodology for this investigation.  To implement this method, we transform the field equation derived from the $f(R)$ theory within the context of an FRW spacetime background containing a perfect fluid into a system of autonomous equations. 
These equations are formulated using appropriately selected dimensionless variables defined in terms of cosmological quantities. 
The fixed points of the set of autonomous equations   represent critical states or scenarios in the evolution of the universe.
Analyzing the stability and properties of these fixed points is essential for understanding the past, present, and future of the cosmos as it  determines whether they correspond to attractors or repellers in the universe's evolution. Stable fixed points represent the long-term behavior of the universe, while unstable fixed points correspond to transient phases.  
If the universe has a stable fixed point corresponding to eternal expansion driven by a $f(R)$ model, it implies that 
according to the model the universe will continue to expand indefinitely.  \\

The stability analysis for the model we've examined reveals four fixed points: $P_1$, $P_2$, $P_3$, and $P_4$, as shown in Table 1. Among these, $P_1$ and $P_2$ are
independent of the model parameter $n$,  while $P_3$ and $P_4$ depend on the value of $n$.
Only $P_2$ proves to be asymptotically stable for all values of $n$ within the range $0 < n < 1$, and it gives $\Omega_{\rm GC} = 1$ (with $\Omega_m = 0$) and $\omega_{\rm eff} = -1$. 
This signifies that $P_2$ serves as an attractor for cosmic acceleration, driven by the geometric curvature term arising from the $f(R)$ modification.
$P_1$, however, remains a saddle point for any $n$ within the range $0 < n < 1$. This point corresponds to an unphysical cosmological solution, as the equation of state parameter $\omega_{\rm eff}$ is positive, whereas the corresponding 
value $\Omega_{\rm GC} = 1$ (with $\Omega_m = 0$) implies acceleration, which necessitates 
$\omega_{\rm eff} < -1/3$.
$P_3$ demonstrates asymptotic stability when $n$ falls within the range of $[0.33 - 0.71]$ but fails to satisfy the constraint that the density parameter $\Omega_m$ remains a positive fraction within this specified range of $n$.
Finally, $P_4$ represents a phantom-like solution for cosmic acceleration within the range $0.5 < n \leqslant 1$, as $\omega_{\rm eff} < -1$ within this range. However, $P_4$ exhibits stability only for the interval $0 < n < 0.28$ which does not overlap with the previously mentioned range of $n$ corresponding to the phantom solution. 
In summary, the nature of the stability of point $P_2$ indicates that there is only one asymptotically stable solution within the considered model. This solution can account for the present-day acceleration of the universe.\\

\textbf{Acknowledgement:} I thank Rahul Roy, Anirban Chatterjee
and Abhijit Bandyopadhyay for useful discussions.
I extend my appreciation to the Department of Physics at RKMVERI for generously providing the essential departmental resources necessary for my work. This study
is a part of my project work within my ongoing Master's program in Physics at the Department of Physics, RKMVERI, India.

\end{document}